\documentclass{article}
\usepackage[T1]{fontenc} % add special characters (e.g., umlaute)
\usepackage[utf8]{inputenc} % set utf-8 as default input encoding
\usepackage{ismir,amsmath,cite,url}
\usepackage{graphicx}
\usepackage{times}
\usepackage{soul}
\usepackage{color}
\usepackage{amsthm}
\usepackage{booktabs}
\usepackage{algorithm}
\usepackage{algorithmic}
\usepackage{amssymb}
\usepackage{enumitem}
\title{Interpreting Song Lyrics with an Audio-Informed Pre-trained Language Model}

\multauthor
{Yixiao Zhang$^1$ \hspace{1cm} Junyan Jiang$^{2,3}$ \hspace{1cm} Gus Xia$^{2,3}$} { \bfseries{Simon Dixon$^1$}\\
 $^1$ Centre for Digital Music, Queen Mary University of London\\
$^2$ Music X Lab, NYU Shanghai \hspace{1cm} $^3$ MBZUAI \\
{\tt\small yixiao.zhang@qmul.ac.uk, jj2731@nyu.edu, gxia@nyu.edu, s.e.dixon@qmul.ac.uk}
}

\def\authorname{Yixiao Zhang, Junyan Jiang, Gus Xia and Simon Dixon}

\begin{document}

\maketitle
\begin{abstract}
% Short texts that summarise and interpret?
Lyric interpretations can help people understand songs and their lyrics quickly, and can also make it easier to manage, retrieve and discover songs efficiently from the growing mass of music archives. In this paper we propose BART-fusion, a novel model for generating lyric interpretations from lyrics and music audio that combines a large-scale pre-trained language model with an audio encoder. We employ a cross-modal attention module to incorporate the audio representation into the lyrics representation to help the pre-trained language model understand the song from an audio perspective, while preserving the language model's original generative performance. We also release the Song Interpretation Dataset, a new large-scale dataset for training and evaluating our model. Experimental results show that the additional audio information helps our model to understand words and music better, and to generate precise and fluent interpretations. An additional experiment on cross-modal music retrieval shows that interpretations generated by BART-fusion can also help people retrieve music more accurately than with the original BART.\footnote{Open-sourced code and pretrained models: \url{https://github.com/ldzhangyx/BART-fusion.}}

\end{abstract}

\section{Introduction} \label{sec:intro}

Lyrics play a key role in the understanding and creation of songs, expressing emotions and delivering messages in the form of natural language \cite{watanabe2020lyrics}. Lyrics have both linguistic and musical characteristics: the field of Lyric Information Processing (LIP) can consequently be seen as a bridge between Music Information Retrieval (MIR) and Natural Language Processing (NLP), encompassing a range of new challenges such as lyric structure analysis \cite{fell2018lyrics}, lyric semantic analysis \cite{delbouys2018music}, automatic lyric generation \cite{sheng2020songmass}, and lyric understanding \cite{tsaptsinos2017lyrics}. In this paper, we focus on the task of multimodal lyric interpretation, which requires the model to understand both the words and music of a song, and to produce a natural, concise and human-like description of its lyrics.

\begin{figure}[t]
    \centering
    \includegraphics[width=\linewidth]{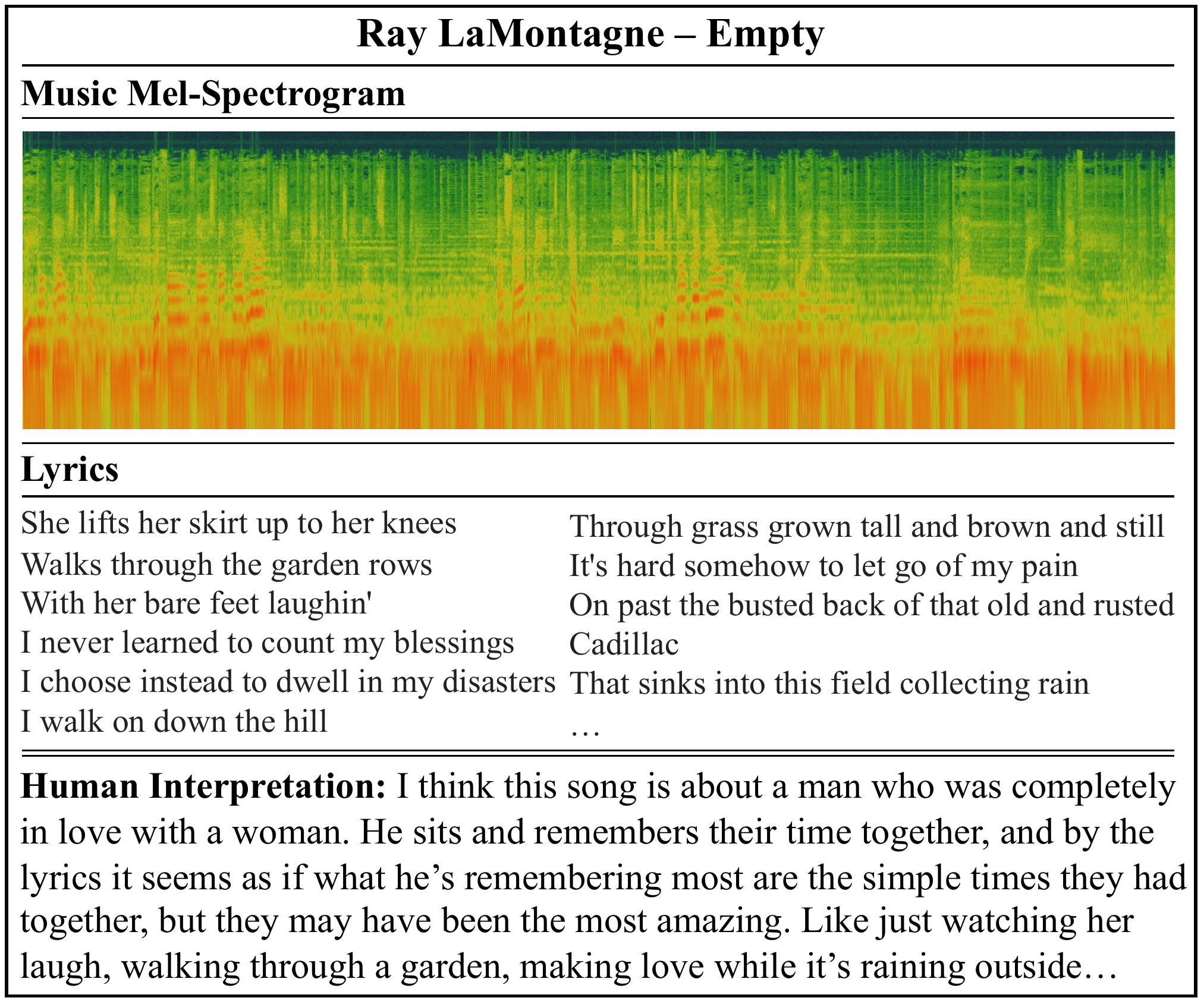}
    \caption{An example of lyrics and their interpretation in real-life, where the interpretation is written by a human. Information from the audio modality includes the representations of instruments, styles, chords, etc., which may help the model to understand the meaning of the lyrics.}
    \label{fig:my_label}
\end{figure}

In real life, human interpretations of lyrics often contain both a general summary of the lyrical theme and a detailed analysis in relation to specific lines. Figure \ref{fig:my_label} shows such an example. Considering that the human interpretation of the lyrics contains subjective elements, the lyrics interpretation task is like an extension of the lyrics summarisation task. The task requires the model to be able to (1) select an excerpt from the lyrics, as in extractive text summarisation; (2) generate explanatory text from lyrics, which is similar to abstractive text summarisation.  However, previous summarisation methods for general texts \cite{al2018hierarchical,lewis2019bart} are not necessarily applicable in the context of lyrics, because song lyrics often contain rich metaphors, poetic themes, and a high degree of rhythm \cite{fell2018lyrics}. Previous studies have attempted to apply extractive summarisation to song lyrics using TextRank algorithms \cite{son2018music} and audio-text alignment algorithms \cite{fell2019song}. The drawback of such approaches is that the summary by itself is not enough to explain the lyrics. To the best of our knowledge, this paper is the first study to use both extractive and abstractive methods to generate lyric interpretations.

Compared to unimodal lyric interpretation models, multimodal models can use information from the music audio domain, such as style, emotion and instrument representations, to reduce the difficulty of understanding lyrics and to improve the quality of the generated text. In recent years, Transformer-based multimodal generative pre-training models have performed well for tasks such as text understanding \cite{li2020unimo, zhang2020butter} and text generation \cite{zhou2020unified, chen2021visualgpt}. Some related works have attempted to adapt existing pre-trained language models to multimodal tasks \cite{yu2021vision} or conditional generation tasks \cite{dathathri2019plug, chen2021visualgpt}. The model we introduce in this paper is inspired by Yu \textit{et al.} \cite{yu2021vision} and we choose to adapt it from a pre-trained language model (BART) \cite{lewis2019bart}. Our model makes use of two modalities: the text of the song lyrics and the corresponding music audio. We add a convolutional encoder (CNN-SA) \cite{won2019toward} to extract a representation of the audio and transfer it to the text domain by computing the similarity between the semantics of the song lyrics and the audio representation through a cross-modal attention mechanism, implemented by a multi-headed attention layer \cite{tsai2019multimodal}. The transformed music audio representation is then fused into the semantic representation of the lyrics as an additional embedding. We discuss more details of the model in Section \ref{sec:method}. To train and evaluate our model, we propose a new dataset, the Song Interpretation Dataset, which contains 27,834 songs with ~490,000 corresponding user interpretations. This is the first large-scale open-source dataset for lyric interpretation. We describe the dataset in detail in Section \ref{sec:dataset}.

We evaluate our model against the original BART model (as a baseline) on the Song Interpretation Dataset, as described in Section \ref{sec:expsetting}. On the lyric interpretation task, our model outperforms the baseline model according to the standard text summarisation metrics ROUGE, METEOR, and BERT-Score. Ablation experiments show that our dataset filtering techniques also improve model performance. To show the value of our model for other tasks, we also present experimental results demonstrating that it performs better than the baseline on a cross-modal retrieval task.

The main contributions of this work can be summarised as follows:

\begin{enumerate}[itemsep=2pt,topsep=0pt,parsep=0pt]
    \item We present BART-fusion, the first multimodal generative model for lyric interpretation. We investigate the integration of audio representations with lyric representations and show that audio representations can improve the performance of lyric interpretation models;
    \item We contribute a large-scale multimodal dataset containing paired audio, lyrics, and lyric interpretations that can be used for music understanding tasks such as lyric interpretation.
\end{enumerate}

% \section{Related Work}

% \subsection{Lyrics Information Processing}

% Lyrics Information Processing (LIP) is a research area that focuses on lyric texts. However, research on lyrics in the MIR community is relatively limited compared to research on audio signals or musical scores, and there has been little related work in the field of NLP as well.The field of LIP can be categorised into three main parts: (1) Lyrics Analysis, such as Rhythm Scheme Identification, Lyrics Segmentation Lyrics Semantic Analysis, Topic Modeling, etc.; (2) Lyrics Generation, such as Melody-Conditioned Lyrics Generation, Rhyme-scheme-Conditioned Lyrics Generation, Ghostwriting, etc.; (3) Applications, such as Content-based Lyrics Exploration and Lyric summarisation.

% \subsection{Multimodal Abstractive summarisation}

\section{Method} \label{sec:method}

In this section, we first revisit the original BART model in Section \ref{sec:bart}. We then discuss the approach to extract musical features from an audio spectrogram in Section \ref{sec:cnnsa}. Finally, we introduce the music-text representation fusion mechanism in Section \ref{sec:fusion}. We identify text-domain features with the subscript $t$ and music-domain features with $m$.

\subsection{BART Model for the Generative Task}\label{sec:bart}

\begin{figure*}[ht]
    \centering
    \includegraphics[width=0.75\linewidth]{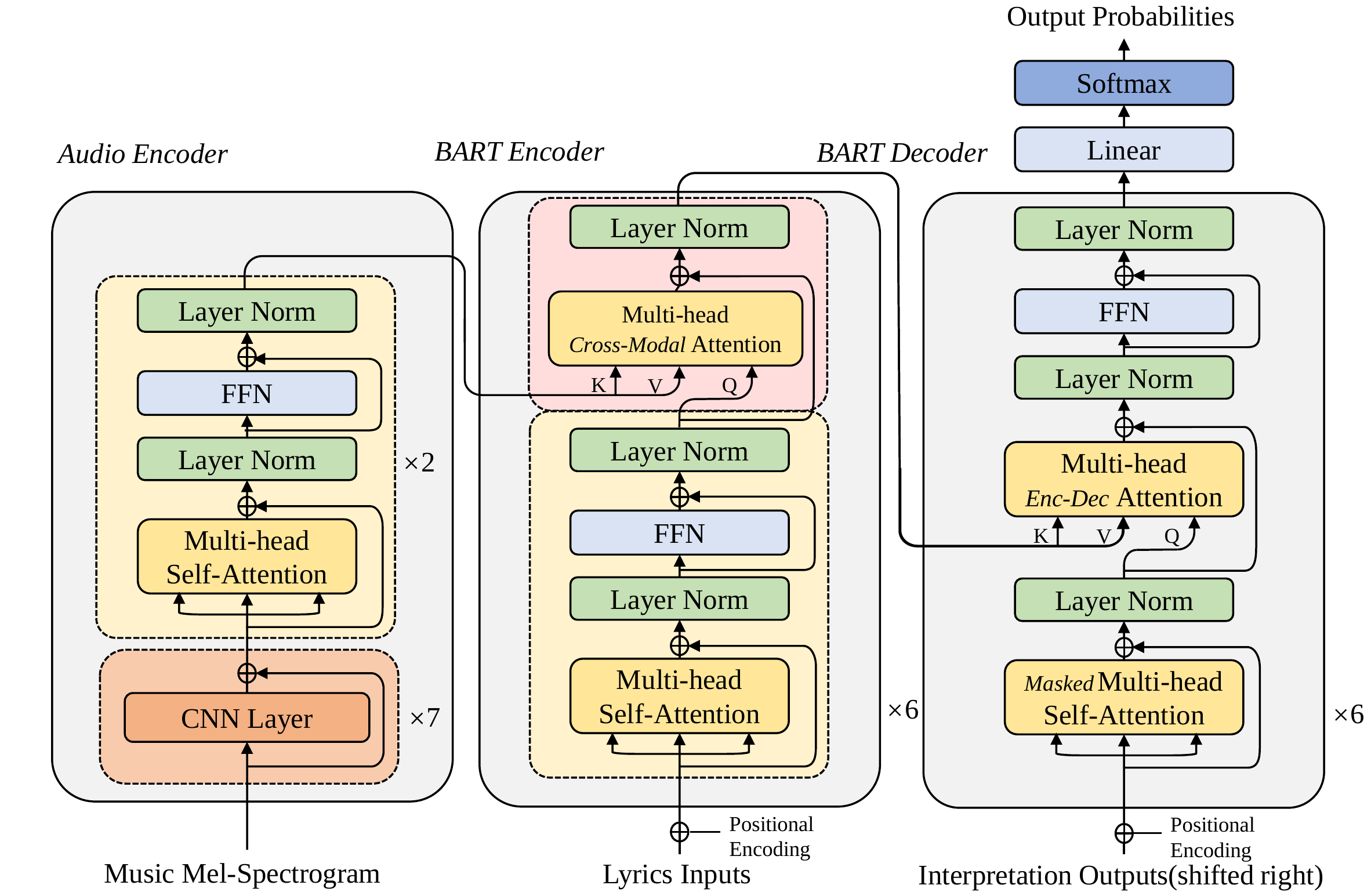}
    \caption{An overview of our proposed model. The model is divided into three parts from left to right: the audio encoder, the BART encoder, and the BART decoder. The fusion of the semantic representation of music audio and lyric text occurs in the upper part of the middle module (pink background). Only at the last two layers of the BART encoder, the music audio representation and the lyric text representation are semantically fused: the music audio representation and the lyric text representation are fed into the cross-modal attention module, and the result is added as an additional embedding to the original lyric text representation. The fused representation is fed into the BART decoder to generate an interpretation. }
    \label{fig:model}
\end{figure*}

Transformer-based pre-trained encoder-decoder language models such as BART \cite{lewis2019bart}, MASS \cite{song2019mass} and T5 \cite{raffel2019exploring}  generalize BERT \cite{devlin2018bert} (due to the bidirectional encoder) and GPT \cite{radford2018improving} (with the left-to-right decoder), achieving good results on sequence-to-sequence tasks such as text summarisation and machine translation. Our model takes advantage of the text generation ability of BART, the structure of which is shown on the right side of Figure \ref{fig:model}.

The lyric text input is firstly tokenized and embedded. We assume the lyric text sequence has $L_t$ tokens, and the embedding dimension is $d_t$, resulting in an embedding $X_t \in \mathbb{R}^{L_t \times d_t}$. Following Vaswani \textit{et al.} \cite{vaswani2017attention}, we add an absolute positional embedding $E_\text{pe}$ to get the final input features $H_t^0$: 

\begin{equation}
    H_{t}^{0} = X_t + E_\text{pe}. \label{eq:position}
\end{equation}

These input features are then passed to the encoder. The encoder has a stack of six layers, as illustrated in Figure \ref{fig:model}, where a single Transformer layer is shown by a yellow box. Each Transformer layer contains a multi-head Self-Attention (SA) module and a Feed-Forward Network (FFN), each followed by a Layer Normalization (LN) module. For the $i$-th layer, the representation is calculated as:

\begin{align}
\widetilde H_{t}^{i} &= \text{LN}(\text{SA}(H_{t}^{i-1}W_Q, H_{t}^{i-1}W_K, H_{t}^{i-1}W_V)W_a\nonumber \\
    &+ H_{t}^{i-1}) \\
    H_{t}^{i} &= \text{LN}(\text{FFN}(\widetilde H_{t}^{i}) + \widetilde H_{t}^{i}),
    \label{eq:attention}
\end{align}
where $H_{t}^{i} \in \mathbb{R}^{L_t \times d_t}$, and $W_Q \in \mathbb{R}^{d_t \times d_{a}}$, $W_K \in \mathbb{R}^{d_t \times d_{a}}$ and $W_V \in \mathbb{R}^{d_t \times d_{a}}$ denote linear transformation matrices which map the representations to a common space. $W_a \in \mathbb{R}^{d_a \times d_{t}}$ linearly projects the attention value back to the desired dimensionality.

The decoder also consists of a stack of six Transformer layers, which is similar to the encoder. But the multi-head self-attention module in the decoder is masked to respect causality, and an additional multi-head encoder-decoder attention is introduced to incorporate the encoder representation.

\subsection{Audio Encoder} \label{sec:cnnsa}

For the multimodal lyric interpretation model, we expect the music audio modality to provide some additional semantic information, such as style, mood, instrumentation, etc., to help the model understand the lyrics better. We design an audio encoder to extract a representation following Won \textit{et al.} \cite{won2019toward}. The audio encoder uses a stack of CNN layers as a filter to extract local features, followed by a self-attention module to capture the global and temporal features of the audio.

The audio encoder receives an audio clip $X_m$ and transforms it to a mel spectrogram $H_{m}^{0}$ as input. We compute the feature map of the $i$-th layer of the encoder as follows:

\begin{align}
    \widetilde H_{m}^{i} &= \text{BN}(\text{CNN}_2(\text{ReLU}(\text{CNN}_1(H_{m}^{i-1})))) \\
    H_{m}^{i} &= \widetilde H_{m}^{i} + \text{BN}(\text{CNN}_3(H_{m}^{i-1})), 
\end{align}
where $\text{BN}$ is the Batch Normalization operation, and the $\text{CNNs}$ are convolutional modules with different parameters. We add a residual connection to each CNN layer. 

We then add two Transformer layers, identical to those in the BART encoder, to extract the final representation of music audio. Finally, we get the music audio representation $Z_m \in \mathbb{R}^{L_{m} \times d_{m}}$, where $L_{m}$ and $d_{m}$ denote the shape of the music audio feature map at the last CNN layer.

\subsection{Representation Fusion}\label{sec:fusion}

As shown in Figure \ref{fig:model}, we insert a representation fusion module into the BART encoder to incorporate musical information. Inspired by Tsai \textit{et al.} \cite{tsai2019multimodal} and Yu \textit{et al.} \cite{yu2021vision}, we apply cross-modal attention to enable the music audio representation to be transferred to the lyric text domain. Jawahar \textit{et al.} \cite{jawahar2019does} have shown that BART encoders tend to extract semantic information in the last few layers, so we only fuse semantic representations at the final two Transformer layers.

For a specific Transformer layer $i$, we have the lyric text representation $H_{t}^{i} \in \mathbb{R}^{L_t \times d_t}$ and the music audio representation $Z_{m} \in \mathbb{R}^{L_{m} \times d_{m}}$, which is the same for each layer. We calculate the domain-adapted music audio representation with a multi-head Cross-Modal Attention (CMA) module:

\begin{equation}
    H_{m\rightarrow t}^{i} = \text{CMA}(H_{t}^{i}W'_Q, Z_{m}W'_K, Z_{m}W'_V)W'_a,
    % H_{m\rightarrow t}^{i} &= \widetilde H_{m\rightarrow t}^{i},
\end{equation}
where $H_{m\rightarrow t}^{i} \in \mathbb{R}^{L_t \times d_t}$ and the symbol ${m\rightarrow t}$ denotes cross-modal attention from music audio to the lyric domain. $d_{a'}$ is the dimensionality of the attention module, and similar to Eq.\ \ref{eq:attention}, $W'_Q \in \mathbb{R}^{d_t \times d_{a'}}$, $W'_K \in \mathbb{R}^{d_m \times d_{a'}}$ and $W'_V \in \mathbb{R}^{d_m \times d_{a'}}$ denote linear transformation matrices which map the representations to a common space. $W'_a \in \mathbb{R}^{d_{a'} \times d_t}$ linearly projects the attention value back to the lyric text dimension.

We add the cross-domain representation as an additional embedding \cite{li2020optimus} to the lyric text representation to get the final representation for the last two layers of the BART encoder:

\begin{equation}
    H_{\text{fusion}}^{i} = H_t^{i} + H_{m\rightarrow t}^{i}.
\end{equation}

\section{Song Interpretation Dataset} \label{sec:dataset}

The lack of suitable datasets has prevented deep learning models from learning to describe songs in natural language. Related studies \cite{choi2016music,muscaps} refer to some datasets, but they share two common problems: 1) the amount of data is small and 2) the datasets are not open-sourced. We propose a new dataset, the Song Interpretation Dataset, for the lyric interpretation task.\footnote{The Song Interpretation Dataset is anonymously open for downloading: \url{https://doi.org/10.5281/zenodo.7019124.}}

The Song Interpretation Dataset combines data from two sources: (1) music and metadata from the Music4All Dataset \cite{santana2020music4all}, and (2) lyrics and user interpretations from SongMeanings.com\footnote{\url{https://songmeanings.com/}}. We design a music metadata-based matching algorithm that aligns matching items in the two datasets with each other. In the end, we successfully match 25.47\% of the tracks in the Music4All Dataset.

The dataset contains audio excerpts from 27,834 songs (30 seconds each, recorded at 44.1 kHz), the corresponding music metadata, about 490,000 user interpretations of the lyric text, and the number of votes given for each of these user interpretations. The average length of the interpretations is 97 words. Music in the dataset covers various genres, of which the top 5 are: Rock (11,626), Pop (6,071), Metal (2,516), Electronic (2,213) and Folk (1,760). 

To the best of our knowledge, this is the first large-scale open-source dataset for lyric interpretation. A comparison with similar datasets is shown in Table \ref{tab:dataset-compare}.

\begin{table}[ht]
    \centering
    \begin{tabular}{l|r|r|c}
    \toprule
     Dataset & Music & Interpretation & Public \\
     \midrule
     Choi \textit{et al.} \cite{choi2016music}    & 800  & 2000 & $\times$ \\
     Manco \textit{et al.} \cite{muscaps} & 17,354 & 17,354 & $\times$ \\
     Ours & \textbf{27,384} & \textbf{490,000} & \checkmark \\
    \bottomrule
    \end{tabular}
    \caption{A comparison of our dataset with previous music description datasets. }
    \label{tab:dataset-compare}
\end{table}

We observe three main issues with interpretations written by real users: (1) some interpretations are very short or very long; (2) interpretations can contain content unrelated to the lyrics themselves, and (3) some interpretations are of low quality. We therefore preprocess the dataset using two techniques:

\begin{enumerate}[itemsep=2pt,topsep=0pt,parsep=0pt]
    \item We \textbf{remove overly short interpretations} with length less than 256 characters to improve data representativeness, since we find that sentences below this length are often meaningless interpretations. For interpretations longer than 2048 characters, we keep only the first 2048 characters, but ensure that the last word is complete.
    \item We use a \textbf{voting-based filtering mechanism} to improve data quality. Every interpretation on SongMeanings.com has a voting result attached, indicating how much the community approves of it, so an interpretation with a higher vote is more likely to be a high-quality interpretation. We therefore create two subsets, keeping only interpretations with positive votes and interpretations with non-negative votes.
\end{enumerate}

To enable the model to be comparable across datasets, we manually select 800 interpretations and use them as a test dataset after excluding them from the original dataset. We have specifically removed all songs that appeared in the test set from the training set to avoid data leakage issues. After the above preprocessing, the dataset has 3 different subsets with the information shown in Table \ref{tab:dataset}.

\begin{table}[ht]
    \centering
    \begin{tabular}{l|r|r|r}
    \toprule
    Dataset Name  & Train & Valid. & Test\\
    \midrule
     Raw dataset    & 440,000 & 50,000 & 800\\
     \midrule
     Dataset Full & 279,283 & 31,032 & 800 \\
     Dataset w/vote $\geq$ 0    & 265,360 & 29,484 & 800 \\
     Dataset w/vote $>$ 0  & 49,736 & 5,526 & 800 \\
    %  \midrule
    %  dataset\_not\_negative\_sum    & 316,574 & 31,574 & 800 \\
    %  dataset\_positive\_sum  & 56,470 & 6,274 & 800 \\
     \bottomrule
    \end{tabular}
    \caption{A comparison of dataset sizes (in number of interpretations) with different filtering methods.}
    \label{tab:dataset}
    % \vspace{-4pt}
\end{table}

\section{Experimental Settings} \label{sec:expsetting}

\begin{table*}[ht]
\centering
\small
\begin{tabular}{l|l|c|ccc|c|c}

\toprule
\textbf{Training dataset}  & \textbf{Method} & \textbf{Data size} & \textbf{R-1} & \textbf{R-2} & \textbf{R-L} & \textbf{METEOR} & \textbf{BERT-Score}\\

   \midrule
 Dataset w/random & BART & 56,470 & 40.0 & 12.5 & 21.7   & 21.1   & \textbf{83.7}   \\
 Dataset w/random & BART-fusion & 56,470 & \textbf{42.1}$^{\ast}$ & \textbf{13.6}$^{\ast}$ & \textbf{23.4}$^{\ast}$    & \textbf{22.0}$^{\ast}$  & {83.3}     \\
 
    \midrule
 Dataset w/voting $>$ 0 & BART & 56,470 & 41.2$_{+}$ & 13.0$_{+}$ & 22.8$_{+}$   & 22.0$_{+}$   & \textbf{83.6}   \\
 Dataset w/voting $>$ 0 & BART-fusion & 56,470 & \textbf{44.3}$^{\ast}_{+}$ & \textbf{14.6}$^{\ast}_{+}$ & \textbf{24.7}$^{\ast}_{+}$    & \textbf{22.6}$^{\ast}_{+}$   & 83.3    \\

\midrule
 Dataset Full        &   BART      &    316,478      & 44.1 & 14.0 & 24.5    & 22.5$_{+}$  & \textbf{83.5} \\
 Dataset Full & BART-fusion     &  316,478    & \textbf{46.1}$^{\ast}$ & \textbf{15.0}$^{\ast}$ & \textbf{25.1}$^{\ast}$   & \textbf{23.0}$^{\ast}$  & \textbf{83.5} \\
 
 \midrule
 Dataset w/voting $\geq$ 0 & BART & 300,712 & 44.8$_{+}$ & {14.9}$_{+}$ & 24.7   & 22.7    & 83.9    \\
 Dataset w/voting $\geq$ 0 & BART-fusion & 300,712 & \textbf{46.7}$^{\ast}_{+}$ & \textbf{15.6}$^{\ast}_{+}$ & \textbf{25.5}$^{\ast}_{+}$  & \textbf{23.4}$^{\ast}$    & \textbf{84.1}    \\

\bottomrule

\end{tabular}
\caption{Evaluation results of BART-fusion and BART (baseline) on the Song Interpretation Dataset with different settings. $\ast$: BART-fusion outperforms BART with $p < 0.05$; $+$: the filtered dataset outperforms the unfiltered dataset with $p < 0.05$.}
\label{tab:booktabs}
\end{table*}

% We denote ROUGE-1, ROUGE-2, ROUGE-L by R-1, R-2 and R-L respectively. 

\subsection{Implementation Details}

% \subsubsection{Data pre-processing} 

We pre-process the lyric text input data by truncating or padding text to 2048 tokens. All the audio signals are downsampled to 16,000 Hz sample rate and converted to short-time Fourier transform representations with a 512-point FFT and 50\%-overlapping Hann window. Finally, we convert these to log mel spectrograms with 128 bins.

% \subsubsection{Hyper-parameters and Optimizer} 

We use \texttt{BART-base} \cite{lewis2019bart} as the pre-trained language model to construct BART-fusion, which has a 6-layer encoder and decoder. For the audio encoder (CNNSA), we use $3 \times 3$ kernels for all layers with [128, 128, 256, 256, 256, 256, 256] channels and [(2, 2), (2, 2), (2, 2), (2, 1), (2, 1), (2, 1), (2, 1)] strides. The cross-modal attention module has 1 head and 1 layer, with $d_a=768$. 

% \subsubsection{Optimizer}

We use AdaFactor \cite{shazeer2018adafactor} as the optimizer. We set the learning rate to $6 \times 10^{-4}$ and reduce it to $6 \times 10^{-5}$ from the 11th epoch. For all experiments, we use a batch size of 8. We train all models for 20 epochs with early stopping of 3 epochs using the ROUGE-1 score on the validation set.

% \subsubsection{Software and hardware.} We use the deep learning framework PyTorch 1.9 to implement our code. We use a single NVIDIA RTX A5000 for all of our experiments.

\subsection{Evaluation Metrics}

Since there are no existing metrics for the lyric interpretation task, we borrow several complementary metrics from similar tasks to evaluate the performance level of the model in a comprehensive manner. We use ROUGE, METEOR and BERT-Score to evaluate the generated interpretations. ROUGE is considered as the main metric for evaluation, because our task is closest to the text summarisation task.

\subsubsection{ROUGE-\{1, 2, L\}} 
ROUGE \cite{lin2004rouge} is a common metric for evaluating abstractive text summaries. It calculates the overlap of 1-gram phrases (R-1), 2-gram phrases (R-2) and their weighted results (R-L). In the context of lyrics, a higher ROUGE score indicates that the generated explanatory text leaves out less necessary information, which is better. 

% \subsubsection{BLEU} BLEU \cite{papineni2002bleu} is a metric used to evaluate machine translation and summarisation models. It measures the precision of model-generated sentences on N-gram phrases. By convention, we follow the same tokenization to WMT'14 and calculate the corpus score, which is the average of 1 to 4-gram results, as the final result.

\subsubsection{METEOR} 
METEOR \cite{banerjee2005meteor} takes into account similar semantic information such as synonyms through WordNet and calculates the similarity score based on F-measure. It complements ROUGE by jointly measuring how semantically similar the model-generated lyric interpretation is to the reference text.

% \subsubsection{SBERT} 
% Sentence-BERT \cite{reimers2019sentence} measures the semantic similarity between the generated and reference sentences. Unlike other metrics, SBERT does not require the model to generate exactly the same words. It is added as an auxiliary metric.

\subsubsection{BERT-Score} 
BERT-Score \cite{zhang2019bertscore} is mainly used to evaluate the naturalness and fluency of the generated text. We expect that incorporating the audio modality should not sacrifice the generative performance of the language model.

We use \texttt{rouge}\footnote{\url{https://github.com/pltrdy/rouge}}, \texttt{nltk}\footnote{\url{https://github.com/nltk/nltk}} and \texttt{bert-score}\footnote{\url{https://github.com/Tiiiger/bert_score}} respectively to compute these metrics.

% \subsection{Baselines}

% \paragraph{Raw Lyrics,} where the original lyrics text are regarded as interpretations. This baseline can be regarded as the upperbound of all extractive methods, because ROUGE is a recall metric, and raw lyrics always recall all lyrics texts excerpted by human references and generated interpretations.

% \paragraph{BART.} In order to evaluate the performance of the multimodal model BERT-fusion, we involve the unimodal BART as a baseline as well.

\section{Results and Analysis}

\subsection{Main Results}

% TODO

We train BART-fusion and the corresponding original BART on several different dataset settings. Results are shown in Table \ref{tab:booktabs}, where all values are the means of multiple independent repeated experiments. We perform a \textit{paired Student's t-test} on the results of BART-fusion and BART, and the results on the filtered datasets and unfiltered datasets respectively, where p-value is 0.05.

Our first finding is that applying a voting-based filtering mechanism to the dataset significantly improves the performance of the models (both BART-fusion and baseline) on this task. For fairness, we take a random subset of \textit{Dataset Full}, which we call \textit{Dataset w/random}, to match the size of \textit{Dataset w/voting $> 0$}. (We still use \textit{Dataset Full} to compare \textit{Dataset w/voting $\geq 0$}.) The experimental results show that the performances of both BART-fusion and the baseline are significantly improved on the filtered datasets.

The experimental results also show that BART-fusion significantly outperforms the baseline, and generates more precise interpretion text for lyrics. For all dataset settings, our BART-fusion models  show better performance on the ROUGE and METEOR scores.

Finally, we find that BART-fusion preserves the generative performance of the pre-trained language model while improving the generation accuracy. The performance of BART-fusion is essentially equal to that of baseline on the BERT-Score metric, which is the metric of text quality. It means that the introduction of audio modal information affects the model mainly semantically and does not affect the naturalness or fluency.

\subsection{Case Study and Error Analysis}

We observe that adding music modality information usually brings the benefits of accurate understanding of the theme, selecting highlighted lyric lines, and emotive sentences from the generated samples. In the case study, we select a representative example showing the generation results from different models with the same lyric input, as shown in Table \ref{tab:examples}. In the first lines, BART-fusion explains the theme of the lyrics more accurately than BART. Then, BART-fusion selects the highlighted lyric lines and gives further detailed interpretation, while the text generated by BART lacks a clear explanation of them. We find that BART-fusion talks about the mood of the song at the end, which is not present in the original BART example.\footnote{We provide a demo page for more examples: \url{https://sites.google.com/view/bart-fusion-ismir2022}.}

We have noticed that when the lyrics are about complex topics, such as religious and philosophical topics, BART-fusion and BART sometimes fail to understand the meaning correctly and generate text that is only superficially correct, which we interpret as a lack of common sense often observed in this class of deep learning models. 

\subsection{Cross-Modal Retrieval Analysis}

\begin{table*}[t]
    \centering
    \small
    \hspace*{-0.4cm}
    \begin{tabular}{p{\linewidth}}
    \toprule
    \textbf{Lyrics:} I cannot find a way to describe it / It's there inside, all I do is hide / I wish that it would just go away / What would you do, you do, if you knew? / What would you do? /  / All the pain I thought I knew / All the thoughts lead back to you / Back to what was never said / Back and forth inside my head / I can't handle this confusion / I'm unable, come and take me away /  / I feel like I am all alone / All by myself, I need to get around this / My words are cold, I don't want them to hurt you / If I show you, I don't think you'd understand / 'Cause no one understands /  / All the pain I thought I knew / All the thoughts lead back to you / Back to what was never said / Back and forth inside my head / I can't handle this confusion / I'm unable, come and take me away / ... \\
    \midrule
    \textbf{Reference:} \textcolor{magenta}{I think this song is about a crush that the singer is afraid to tell the person they are in love with}. She wonder if maybe the other person feels the same, but is also too afriad to tell her. Thinking "what if". Since, she won't admit it she'll never know and this not knowing ends up eating her up inside. Each day she wonders if she could be with this person who she really loves, but she just gets frustrated and eventually wishes that she never even fell in love with the person since it would make her life so much easier. It builds up so much that she just wishes someone would \textcolor{blue}{"come and take me away"}. That could mean two things. It could be death/suicide because she decides it's not worth it or she could just wish for someone else to date so she can just say she's taken and not have to worry about this crush.   \\
    \midrule
    \textbf{BART:} \textcolor{magenta}{I think this song is about a girl who is in love with a guy, but she doesn't know how to tell him because she's afraid he'll hurt her}. She's trying to convince herself that he loves her, but he doesn't understand how she feels about him. She wants him to come and take her away from all the pain she's caused and make her feel better about herself. This is what I think of when I listen to this song. I think it's about someone who is trying to tell someone they love them, but they don't want to tell them because they're afraid that they'll hurt them. I love this song, it's one of my favorites on the album. \\
    \midrule
    \textbf{BART-fusion:} \textcolor{magenta}{I think this song is about a girl who is in love with a guy but she is afraid to tell him because she doesn't want to ruin the friendship}. {In the first verse} she is talking about how she can 't find a way to describe how she feels and how she wishes it would just go away. {The second verse} is about how the guy is telling her how he feels but she isn't sure if he really feels the same way. {The chorus is} her saying how she is confused about what to say and how to tell the guy. \textcolor{blue}{"I'm unable, come and take me away"} means that she is unable to tell her feelings and she is trying to get the guy out of her head. {The last verse} is her telling the guy that she feels like she is going nowhere in her life, and she just wants him to take her away from all of this. That's just my take on it. I can totally relate to this song. \textcolor{red}{It's a very powerful song.}\\
    \bottomrule
    \end{tabular}
    \caption{A generation example for Avril Lavigne - \textit{Take Me Away}.}
    \label{tab:examples}
    % \vspace{-6pt}
\end{table*}

In addition, we design a cross-modal retrieval experiment to test the retrieval capability of our models. We randomly extract a sentence from the real interpretation in the test set as a query string, use this to search over a database of lyric text and musical audio on a semantic level, and return a ranked list of all possible results. We expect the correct result corresponding to the query to be as close to the top of this ranking as possible.

To accomplish this task, the model generates interpretations for all song lyrics in the database, calculates their semantic features using Sentence-BERT \cite{reimers2019sentence}, and stores features in advance. At query time, we also use Sentence-BERT to compute the semantic feature for the query string. We return the song whose feature is most similar to the query feature by computing the cosine similarities between the query feature and features in the database.

We use Mean Reciprocal Rank (MRR) \cite{chapelle2009expected}, a common metric for information retrieval tasks, to measure the performance of the models on this task. Formally, for a set of query strings $Q=\{q_1, q_2, ... , q_m\}$ and the corresponding database $S = \{s_1, s_2, ... , s_n\}$, the model outputs a list of songs sorted by probability for the $i$-th query, where the correct song is ranked $k_i$-th, and the model is scored as:

\begin{equation}
    \text{score}_i = \frac{1}{k_i}, \text{where } k_i \leq n.
\end{equation}

\noindent The final MRR is calculated by averaging the scores:

\begin{equation}
    \text{MRR} = \frac{1}{m}\sum^m_{i=1}\text{score}_i = \frac{1}{m}\sum^m_{i=1}\frac{1}{k_i}.
\end{equation}

From Table \ref{tab:mir}, we find that BART-fusion outperforms the original BART on all 4 dataset settings, i.e., the interpretations generated by BART-fusion can help users find music more accurately. If we return a random ranking result, the MRR value is about 0.9\%, which is much lower than the model performance.

\begin{table}[ht]
\centering
\small
\begin{tabular}{l|l|c}
\toprule
\textbf{Dataset}  & \textbf{Method} & \textbf{MRR (\%)}  \\
\midrule
Dataset w/random         &   BART   & 25.3 \\
Dataset w/random    & BART-fusion &  \textbf{27.5}$^{\ast}$   \\
\midrule
Dataset w/voting $>$ 0         &   BART   & 26.1$_{+}$  \\
Dataset w/voting $>$ 0    & BART-fusion & \textbf{28.2}$^{\ast}_{+}$    \\
\midrule
Dataset Full         &   BART       & 26.2  \\
Dataset Full    & BART-fusion & \textbf{30.5}$^{\ast}$    \\
\midrule
Dataset w/voting $\geq$ 0         &   BART   & 26.4  \\
Dataset w/voting $\geq$ 0    & BART-fusion & \textbf{32.5}$^{\ast}_{+}$    \\
\bottomrule
\end{tabular}
\caption{Evaluation results for the music retrieval task. $\ast$: BART-fusion outperforms BART with $p < 0.05$; $+$: the filtered dataset outperforms the unfiltered dataset with $p < 0.05$.}
\label{tab:mir}
\vspace{-5pt}
\end{table}

% \vspace{-5pt}
\section{Conclusion and Future Work}

In this paper, we have proposed a novel model to generate interpretations from song lyrics and musical audio. Our modelis based on a pre-trained language model and generates better lyric interpretations by fusing text and music semantic representations. We also have proposed a new dataset and explored the process of dataset creation, and have investigated how different treatments of the dataset can affect the performance of the model. We have designed an additional experiment that shows that our model outperforms BART on cross-modal music retrieval tasks. Our work has a range of potential applications, such as helping people better understand English lyrics (especially for non-native English speakers) and natural language based music discovery.

The current model still has shortcomings, such as difficulty in understanding complex topics and lack of general common sense. In future work, we will try to improve the model by adding metadata and knowledge base inputs.  We also plan to extend the model to describe the musical content of a song in natural language. In addition, large-scale language models may be biased, reinforce stereotypes and introduce harms, which deserves our attention in the future.

\section{Acknowledgement}

We want to thank Mark Levy for his great contribution to this work. We also thank Yin-Jyun Luo and Liming Kuang for their enthusiastic help during the writing process of the paper. Yixiao Zhang is a research student at the UKRI Centre for Doctoral Training in Artificial Intelligence and Music, supported jointly by the China Scholarship Council, Queen Mary University of London and Apple Inc.

% For bibtex users:
\bibliography{ISMIR}

\begin{thebibliography}{10}
\providecommand{\url}[1]{#1}
\csname url@samestyle\endcsname
\providecommand{\newblock}{\relax}
\providecommand{\bibinfo}[2]{#2}
\providecommand{\BIBentrySTDinterwordspacing}{\spaceskip=0pt\relax}
\providecommand{\BIBentryALTinterwordstretchfactor}{4}
\providecommand{\BIBentryALTinterwordspacing}{\spaceskip=\fontdimen2\font plus
\BIBentryALTinterwordstretchfactor\fontdimen3\font minus
  \fontdimen4\font\relax}
\providecommand{\BIBforeignlanguage}[2]{{%
\expandafter\ifx\csname l@#1\endcsname\relax
\typeout{** WARNING: IEEEtran.bst: No hyphenation pattern has been}%
\typeout{** loaded for the language `#1'. Using the pattern for}%
\typeout{** the default language instead.}%
\else
\language=\csname l@#1\endcsname
\fi
#2}}
\providecommand{\BIBdecl}{\relax}
\BIBdecl

\bibitem{watanabe2020lyrics}
K.~Watanabe and M.~Goto, ``Lyrics information processing: Analysis, generation,
  and applications,'' in \emph{Proceedings of the 1st Workshop on NLP for Music
  and Audio (NLP4MusA)}, 2020, pp. 6--12.

\bibitem{fell2018lyrics}
M.~Fell, Y.~Nechaev, E.~Cabrio, and F.~Gandon, ``Lyrics segmentation: Textual
  macrostructure detection using convolutions,'' in \emph{COLING}, 2018, pp.
  2044--2054.

\bibitem{delbouys2018music}
R.~Delbouys, R.~Hennequin, F.~Piccoli, J.~Royo-Letelier, and M.~Moussallam,
  ``Music mood detection based on audio and lyrics with deep neural net,''
  \emph{arXiv preprint arXiv:1809.07276}, 2018.

\bibitem{sheng2020songmass}
Z.~Sheng, K.~Song, X.~Tan, Y.~Ren, W.~Ye, S.~Zhang, and T.~Qin, ``Songmass:
  Automatic song writing with pre-training and alignment constraint,''
  \emph{arXiv preprint arXiv:2012.05168}, 2020.

\bibitem{tsaptsinos2017lyrics}
A.~Tsaptsinos, ``Lyrics-based music genre classification using a hierarchical
  attention network,'' \emph{arXiv preprint arXiv:1707.04678}, 2017.

\bibitem{al2018hierarchical}
K.~Al-Sabahi, Z.~Zuping, and M.~Nadher, ``A hierarchical structured
  self-attentive model for extractive document summarization ({HSSAS}),''
  \emph{IEEE Access}, vol.~6, pp. 24\,205--24\,212, 2018.

\bibitem{lewis2019bart}
M.~Lewis, Y.~Liu, N.~Goyal, M.~Ghazvininejad, A.~Mohamed, O.~Levy, V.~Stoyanov,
  and L.~Zettlemoyer, ``{BART}: Denoising sequence-to-sequence pre-training for
  natural language generation, translation, and comprehension,'' \emph{arXiv
  preprint arXiv:1910.13461}, 2019.

\bibitem{son2018music}
J.~Son and Y.~Shin, ``Music lyrics summarization method using textrank
  algorithm,'' \emph{Journal of Korea Multimedia Society}, vol.~21, no.~1, pp.
  45--50, 2018.

\bibitem{fell2019song}
M.~Fell, E.~Cabrio, F.~Gandon, and A.~Giboin, ``Song lyrics summarization
  inspired by audio thumbnailing,'' in \emph{RANLP}, 2019.

\bibitem{li2020unimo}
W.~Li, C.~Gao, G.~Niu, X.~Xiao, H.~Liu, J.~Liu, H.~Wu, and H.~Wang, ``{UNIMO}:
  Towards unified-modal understanding and generation via cross-modal
  contrastive learning,'' \emph{arXiv preprint arXiv:2012.15409}, 2020.

\bibitem{zhang2020butter}
Y.~Zhang, Z.~Wang, D.~Wang, and G.~Xia, ``{BUTTER}: A representation learning
  framework for bi-directional music-sentence retrieval and generation,'' in
  \emph{Proceedings of the 1st Workshop on NLP for Music and Audio (NLP4MusA)},
  2020, pp. 54--58.

\bibitem{zhou2020unified}
L.~Zhou, H.~Palangi, L.~Zhang, H.~Hu, J.~Corso, and J.~Gao, ``Unified
  vision-language pre-training for image captioning and {VQA},'' in
  \emph{AAAI}, vol.~34, no.~07, 2020, pp. 13\,041--13\,049.

\bibitem{chen2021visualgpt}
J.~Chen, H.~Guo, K.~Yi, B.~Li, and M.~Elhoseiny, ``{VisualGPT}: Data-efficient
  adaptation of pretrained language models for image captioning,'' \emph{arXiv
  preprint arXiv:2102.10407}, 2021.

\bibitem{yu2021vision}
T.~Yu, W.~Dai, Z.~Liu, and P.~Fung, ``Vision guided generative pre-trained
  language models for multimodal abstractive summarization,'' \emph{arXiv
  preprint arXiv:2109.02401}, 2021.

\bibitem{dathathri2019plug}
S.~Dathathri, A.~Madotto, J.~Lan, J.~Hung, E.~Frank, P.~Molino, J.~Yosinski,
  and R.~Liu, ``Plug and play language models: A simple approach to controlled
  text generation,'' \emph{arXiv preprint arXiv:1912.02164}, 2019.

\bibitem{won2019toward}
M.~Won, S.~Chun, and X.~Serra, ``Toward interpretable music tagging with
  self-attention,'' \emph{arXiv preprint arXiv:1906.04972}, 2019.

\bibitem{tsai2019multimodal}
Y.-H.~H. Tsai, S.~Bai, P.~P. Liang, J.~Z. Kolter, L.-P. Morency, and
  R.~Salakhutdinov, ``Multimodal transformer for unaligned multimodal language
  sequences,'' in \emph{ACL}, vol. 2019.\hskip 1em plus 0.5em minus 0.4em\relax
  NIH Public Access, 2019, p. 6558.

\bibitem{song2019mass}
K.~Song, X.~Tan, T.~Qin, J.~Lu, and T.-Y. Liu, ``{MASS}: Masked sequence to
  sequence pre-training for language generation,'' \emph{arXiv preprint
  arXiv:1905.02450}, 2019.

\bibitem{raffel2019exploring}
C.~Raffel, N.~Shazeer, A.~Roberts, K.~Lee, S.~Narang, M.~Matena, Y.~Zhou,
  W.~Li, and P.~J. Liu, ``Exploring the limits of transfer learning with a
  unified text-to-text transformer,'' \emph{arXiv preprint arXiv:1910.10683},
  2019.

\bibitem{devlin2018bert}
J.~Devlin, M.-W. Chang, K.~Lee, and K.~Toutanova, ``{BERT}: Pre-training of
  deep bidirectional transformers for language understanding,'' \emph{arXiv
  preprint arXiv:1810.04805}, 2018.

\bibitem{radford2018improving}
A.~Radford, K.~Narasimhan, T.~Salimans, I.~Sutskever \emph{et~al.}, ``Improving
  language understanding by generative pre-training,'' 2018.

\bibitem{vaswani2017attention}
A.~Vaswani, N.~Shazeer, N.~Parmar, J.~Uszkoreit, L.~Jones, A.~N. Gomez,
  {\L}.~Kaiser, and I.~Polosukhin, ``Attention is all you need,'' in
  \emph{Advances in Neural Information Processing Systems}, 2017, pp.
  5998--6008.

\bibitem{jawahar2019does}
G.~Jawahar, B.~Sagot, and D.~Seddah, ``What does {BERT} learn about the
  structure of language?'' in \emph{ACL}, 2019.

\bibitem{li2020optimus}
C.~Li, X.~Gao, Y.~Li, B.~Peng, X.~Li, Y.~Zhang, and J.~Gao, ``{OPTIMUS}:
  Organizing sentences via pre-trained modeling of a latent space,''
  \emph{arXiv preprint arXiv:2004.04092}, 2020.

\bibitem{choi2016music}
K.~Choi, J.~H. Lee, X.~Hu, and J.~S. Downie, ``Music subject classification
  based on lyrics and user interpretations,'' \emph{AIST}, vol.~53, no.~1, pp.
  1--10, 2016.

\bibitem{muscaps}
\BIBentryALTinterwordspacing
I.~Manco, E.~Benetos, E.~Quinton, and G.~Fazekas, ``Muscaps: Generating
  captions for music audio,'' \emph{CoRR}, vol. abs/2104.11984, 2021. [Online].
  Available: \url{https://arxiv.org/abs/2104.11984}
\BIBentrySTDinterwordspacing

\bibitem{santana2020music4all}
I.~A.~P. Santana, F.~Pinhelli, J.~Donini, L.~Catharin, R.~B. Mangolin, V.~D.
  Feltrim, M.~A. Domingues \emph{et~al.}, ``Music4all: A new music database and
  its applications,'' in \emph{IWSSIP}.\hskip 1em plus 0.5em minus 0.4em\relax
  IEEE, 2020, pp. 399--404.

\bibitem{shazeer2018adafactor}
N.~Shazeer and M.~Stern, ``Adafactor: Adaptive learning rates with sublinear
  memory cost,'' in \emph{ICML}.\hskip 1em plus 0.5em minus 0.4em\relax PMLR,
  2018, pp. 4596--4604.

\bibitem{lin2004rouge}
C.-Y. Lin, ``{ROUGE}: A package for automatic evaluation of summaries,'' in
  \emph{Text Summarization Branches Out: Proceedings of the ACL-04 Workshop},
  2004, pp. 74--81.

\bibitem{banerjee2005meteor}
S.~Banerjee and A.~Lavie, ``{METEOR}: An automatic metric for {MT} evaluation
  with improved correlation with human judgments,'' in \emph{Proceedings of the
  ACL Workshop on Intrinsic and Extrinsic Evaluation Measures for Machine
  Translation and/or Summarization}, 2005, pp. 65--72.

\bibitem{zhang2019bertscore}
T.~Zhang, V.~Kishore, F.~Wu, K.~Q. Weinberger, and Y.~Artzi, ``{BERTScore}:
  Evaluating text generation with {BERT},'' \emph{arXiv preprint
  arXiv:1904.09675}, 2019.

\bibitem{reimers2019sentence}
N.~Reimers and I.~Gurevych, ``Sentence-{BERT}: Sentence embeddings using
  {S}iamese {BERT}-networks,'' \emph{arXiv preprint arXiv:1908.10084}, 2019.

\bibitem{chapelle2009expected}
O.~Chapelle, D.~Metlzer, Y.~Zhang, and P.~Grinspan, ``Expected reciprocal rank
  for graded relevance,'' in \emph{Proceedings of the 18th ACM Conference on
  Information and Knowledge Management}, 2009, pp. 621--630.

\end{thebibliography}

\end{document}